\documentclass[prl, twocolumn]{revtex4-2}
\usepackage{amssymb}
\usepackage{amsmath}
\usepackage{epsfig}
\usepackage{bm}

\begin{document}

\title{Flow bistability in non-Newtonian electron fluid}

\author{A.~N.~Afanasiev and P.~S.~Alekseev}

\affiliation{Ioffe Institute, 194021 St.~Petersburg, Russia}

\begin{abstract}

Modern two dimensional conductors with low defect densities and strong electron-electron scattering are favorable platforms for formation of a viscous fluid of conduction electrons. Electric properties of these systems are determined by the hydrodynamic regime of charge transport distinguished by many experimental signatures: a decrease in sample resistance with increasing temperature (the Ghurzhi effect), strong negative magnetoresistance and others. Here we consider the flow of 2D electron fluid in the nonlinear regime characterized by non-Newtonian viscosity which depends on spatial gradients of hydrodynamic velocity. We derive a simplified version of the dynamic equations for the non-Newtonian electron fluid and consider the specific underlying mechanism associated with local electron heating. Recent works have demonstrated that this may be one of the main mechanisms for nonlinearity in 2D electron fluids. We show that in a certain range of parameters, the two steady-state flow configurations coexist for the narrow channel geometry, and this bistability leads to an S-shaped current-voltage characteristic. By solving the derived time-dependent dynamic equations, we trace the transient response to a step variation of the longitudinal voltage and demonstrate how the current switching and hysteresis occur in samples with the non-Newtonian electron fluid.

\end{abstract}

\maketitle


{\em 1. Introduction.} High-mobility conductors with very low defect densities enable formation of a viscous fluid of conductive electrons which properties are controlled by frequent inter-particle scattering.
The corresponding hydrodynamic regime of electric transport
 was first reliably identified  in high quality samples of layered
metal PdCoO$_2$~\cite{Moll},
 in single-layered graphene~\cite{graphene_1,Polini_Geim,Levitov_et_al},
 and in GaAs quantum wells~[5-14].
In the first two systems this regime was detected by the strong dependence of
the resistance on the sample size and geometry. In the latter systems,
 it was first identified by the strong negative magnetoresitance, induced
by the magnetic field dependence of the electron  viscosity
 \cite{Gusev_1,je_visc,recentest_,exps_neg_1,exps_neg_2,exps_neg_3,exps_neg_4,Gurzhi_Shevchenko},
and  by the dependence of the sample
 resistance on its complex geomentry~\cite{Keser}.

In number of experiments complex two-dimensional flows of
 the electron fluid were studied in samples with an array
of macroscopic obstacles (disks)~\cite{d1,d1_new,d2,d2_new}.
In Refs.~\cite{disks,disks2} a theory
of hydrodynamic magnetotransport  in such systems was developed.
The effects of the transition between hydrodynamic and ballistic
 flow regimes with an increase in the magnetic field
were examined for very pure samples (see, for example,
Refs.~\cite{1,2,Scaffidi2017,a,Holder,c}).
Other various hydrodynamic transport effects were studied
  in Refs.~\cite{Afanasiev2022,Denisov2022,Glazov,Denisov2023,Alekseev_2023,
 Afanasiev_at_al_2025,Alekseev_Dmitriev,Novikov,Levin2024}.

The anomalies in high-frequency nonlinear transport
of 2D electrons in purest samples were observed
in Refs.~\cite{Smet,Dai2010,Hatke2011,Bandurin2022}
(primarily, a strong resonance  in the photoresistance
at a doubled cyclotron frequency and independence of MIRO on helicity
 of the circularly polarized radiation) and puzzled
specialists in this field.  A theory of
high-frequency hydrodynamic  transport was developed in
 works~\cite{vis_res_0,vis_res_1,vis_res_2,Semiconductors,Afanasiev_2023,Afanasiev_at_al_2025}.
 It was demonstrated that these unusual features
can be explained as a formation of high-frequency electron flow and manifestation
of the resonance in the high-frequency viscosity coefficients
of 2D electrons \cite{vis_res_2,new,Afanasiev_at_al_2025}.  Although there
are non-hydrodynamic theories that explain
 a number of high-frequency transport phenomena in 2D electron systems
 within the model of non-interacting electrons  in disordered samples
(oscillations of the MIRO photoresistance,
their temperature and ac-power dependencies)
 \cite{Vavilov2004,S7,Dmitriev_2012,Beltukov_Dyakonov},
taking into account the inter-electron interaction
and the formation of a viscous flows is apparently critical
for the anomalous effects listed above: giant negative magnetoresistance
 \cite{Gusev_1,recentest_,exps_neg_1,exps_neg_2,exps_neg_3,exps_neg_4},
a large  peak in photoresistance  \cite{Dai2010,Hatke2011,Bandurin2022}
as well as independence of MIRO on the sign
 of the circular polarization of radiation \cite{Smet}.

Experimental studies of non-linear dc transport of 2D electrons in ultra-pure samples
 of GaAs quantum wells were performed
 in Refs.~\cite{non-lin_hydr_1,non-lin_hydr_3,non-lin_hydr_4,
 non-lin_hydr_5,non-lin_hydr_6,n-N}.
In particular, it was shown that the giant negative magnetoresistance effect specific for
stationary flows of electron fluid
 in magnetic field changes the shape
with the increase of the current: it becomes non-monotonic
in the region of small fields,
decreases with increasing current
 in the zero magnetic field, and increases with the current at large magnetic
fields~\cite{non-lin_hydr_1,non-lin_hydr_5,non-lin_hydr_6}.
 In Refs.~\cite{new,Alekseev_Semina_2025,Alekseev_Semina_2026}
  the theories of non-linear electron transport
due to the memory effects
in the electron-electron interaction
and the effect of electron heating by current were proposed and developed. They predict, in particular, that the slow enough flow can be described by
the Navier-Stokes equation with non-linear viscosity depending  on the flow velocity gradient, i.e. the electron fluid should be considered as non-Newtonian.
This leads to the non-trivial differential magnetoresistance
of a sample which was apparently   observed in several independent
experiments~\cite{non-lin_hydr_1,non-lin_hydr_5,non-lin_hydr_6}
  on high-quality GaAs quantum wells.   This evidences that 2D
electron non-Newtonian fluid is formed is such systems.

In this Letter we study stationary and non-stationary large-amplitude flows
of non-Newtonian electron fluid in realistic structures.
In the framework of the model of Refs.~\cite{Alekseev_Semina_2025,Alekseev_Semina_2026} we additionally take into account weak momentum-relaxing electron scattering
by rare residual defects in the bulk of the sample. This provides Ohmic contribution to transport of electron fluid which regularizes stationary flows with arbitrary large amplitude.
 We demonstrate that when the electric field magnitude becomes larger
 than some threshold value, two types of steady-state flow -- the small amplitude
one and the high amplitude one -- coexist due to strong non-Newtonian effects.
We show that these flows are stable and it is possible to switch between them by applying the electric field smaller or greater than the threshold values. Thus, we predict
a bistable S-type current-voltage characteristic of a realistic sample
with the non-Newtonian  electron fluid. Finally, we analyze how the current hysteresis occurs in the transient response of non-Newtonian hydrodynamic conductor with this type of I-V characteristic.


{\em 2. Model of non-Newtonian 2D electron fluid. }
As a minimal model to study the non-linear behavior of the 2D non-Newtonian electron fluid,
we consider the quasi one-dimensional (Poiseuille-like) flows  in perpendicular magnetic field $\mathbf{B}$ in long and relatively narrow samples, with
the width $W$ much smaller  than the plasmon wavelength, see Refs.~\cite{vis_res_2,new,Alekseev_Semina_2025,Alekseev_Semina_2026}.
For this geometry, the $y$-component of  the flow velocity, $V_y$, related to an internal electric  field $\mathbf{E}_y ^{int} (y,t) $ and a non-equilibrium charge density~$ e \,\delta n (y,t)$,
is suppressed~\cite{vis_res_2}.  As a result, the motion equations
for the hydrodynamic velocity $ V(y,t)  \equiv V_x (y,t) $ and
 the shear stress~$  \hat{\sigma} (y,t)  = -\hat{\Pi} (y,t) $
take the form~\cite{f2}:
  \begin{equation}
\label{main_eq_gen}
\left\{
\begin{array}{l}
\displaystyle
     \frac{   \partial V   }{ \partial t   } \, = \,
          \frac{e   E }{m} \,  - \, \frac{1 }{m} \,
           \frac{  \partial \Pi_{xy}  }{ \partial y}
          \,  - \, \frac{ V }{\tau_1}
 \:,
\\\\
\displaystyle
 \frac{\partial \Pi_{xx} }{\partial  t }=  2 \omega_c  \Pi_{xy}
   - \frac{ \Pi_{xx} }{ \tau_{2}(y)}
          \: ,
\\\\
\displaystyle
 \frac{\partial \Pi_{xy} }{\partial  t }
 = - \frac{ \Pi_{xy} }{ \tau_{2}(y) }  - 2 \omega_c  \Pi_{xx}  -
    \frac{   m \eta_0 }{ \tau_{2} }
       \: \frac{  \partial V }{ \partial{y} }  \, ,
\end{array}
\right.
\end{equation}
where $\tau_1$ is the electron momentum relaxation time due to
scattering by disorder and/or phonons,
$\omega_c =eB/(mc)$ is the electron cyclotron frequency,
$\eta_0 = v_F^2 \tau_2 /4$ is the electron viscosity in zero magnetic field,
$\tau _2$ is  the shear stress relaxation
 time due to inter-particle collisions at the temperature of the phonon bath~$T_{ph}$,
  $ \tau_2 (y) = \tau_2 [ \partial V /\partial y]$
is the shear stress relaxation time at the local electron temperature $T_e(y)$ which deviates from $T_{ph}$ due to local viscous heating~\cite{Alekseev_Semina_2026};
 the electric field $E \equiv E_x (t)  $ is due  to the applied voltage.

\begin{figure}[t!]
\centerline{\includegraphics[width=.99 \linewidth]{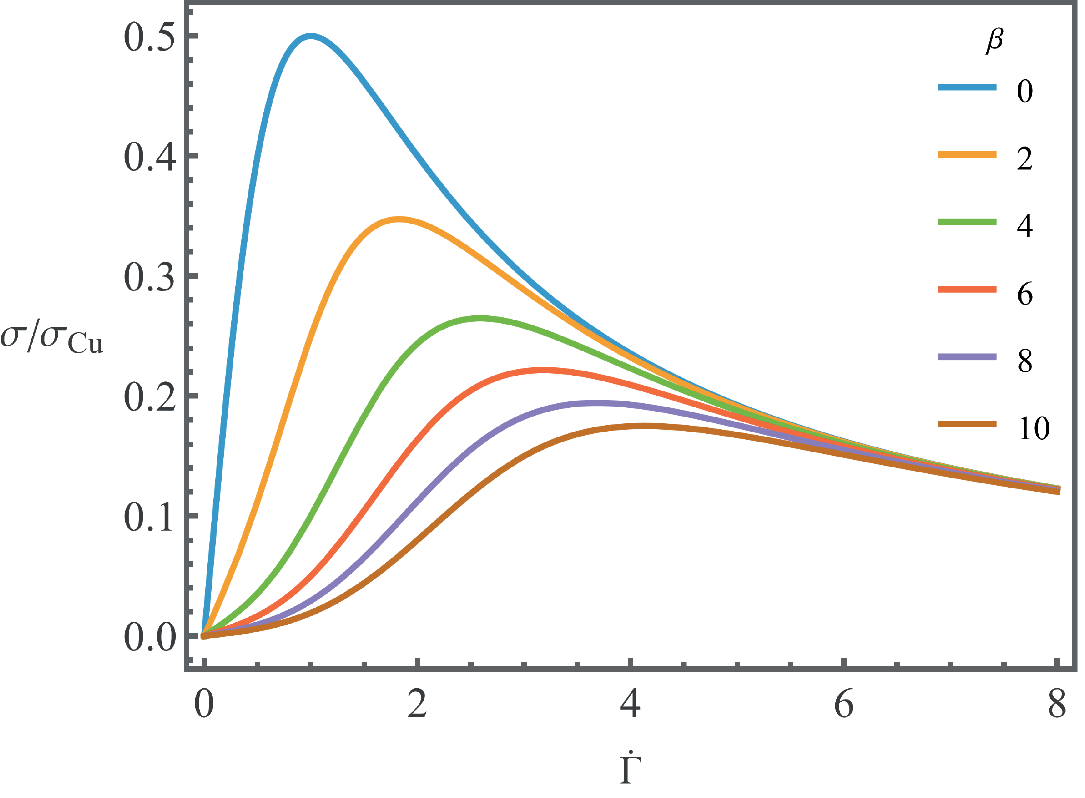}}
\caption{
Dependence of the shear stress component
 $\sigma = \sigma_{xy} =  m\, \eta _{xx}^{nl}  [V] \, \partial V/ \partial y $
 on the velocity gradient
 $\dot{\gamma} = \partial V  /\partial  y$
 for the local heating mechanism of non-Newtonian behavior of 2D electron fluid. The units of $\sigma$ is $\sigma_{\rm Cu} = \eta_0/\sqrt{\tau_{\Delta}\tau_2}$,
  and the dimensionless velocity gradient is $\dot{\Gamma}=\sqrt{\tau_{\Delta}\tau_2}\dot{\gamma}$. Lines of different colors refer to different values of magnetic fields $\beta=2\omega_c \tau_2$.
}
\end{figure}

 The heating effect is controlled by the balance of heat released in the fluid due to
 viscous dissipation $m \eta_{xx} (\partial V /\partial  y)^2 $ per unit time per one electron
and the heat exchange with the phonon system $\alpha (T_e(y) -T_{ph})$~\cite{Karpus_1}.
Here $\eta_{xx} = \eta_0 /[1+(2\omega_c\tau_{2})^2]$  is the diagonal dissipative viscosity in magnetic field.
  This yields the increase of the local electron temperature by $\Delta  T_e (y) = T_e(y)-T_{ph}$ and the local shear relaxation rate $ 1/\tau_2 (y) = 1/\tau_2 + \tau_\Delta (\partial V /\partial  y)^2 $.
 We assume the heating to be relatively small $\Delta  T_e (y) \ll T_{ph}$. The coefficient $\alpha$ and the corresponding
   time $\tau_\Delta (B) \sim \eta_{xx} $ for GaAs quantum wells were analyzed in Ref.~\cite{Alekseev_Semina_2026}.

\begin{figure*}[t!]
\centerline{\includegraphics[width=.99 \linewidth]{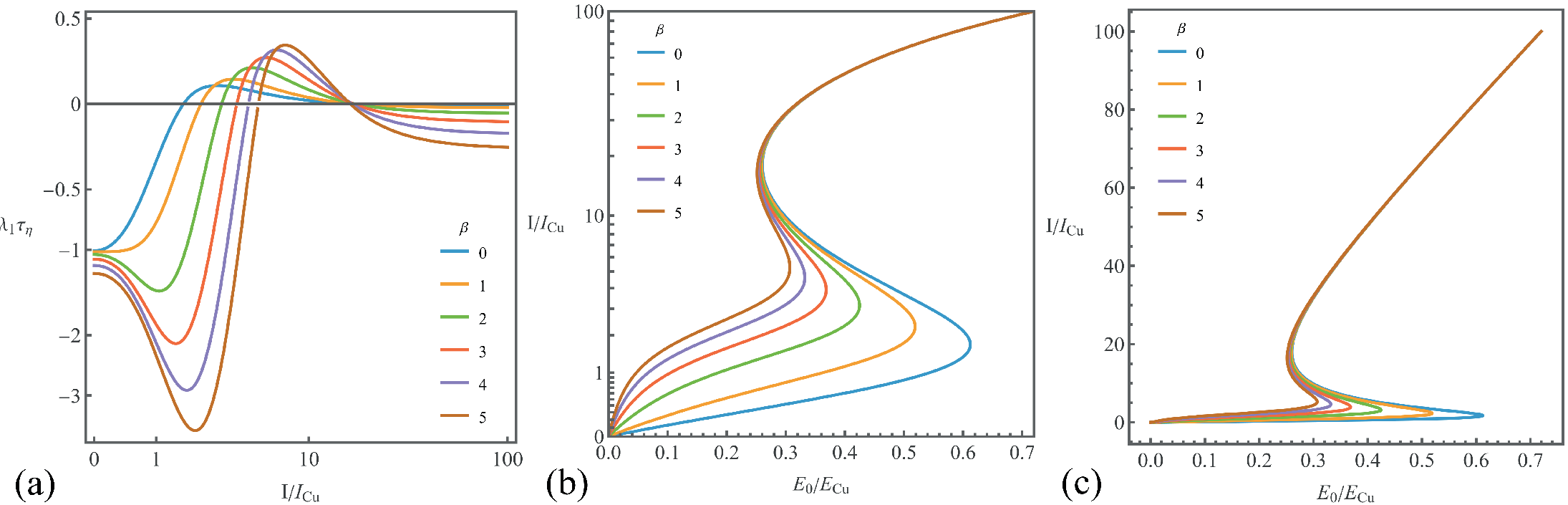}}
\caption{
($a$): Decay/growth rates of the small perturbations $\delta v \, e^{\lambda _1 t}  \ll v $ for the stationary solutions of the equation $\mathcal{F}(v;\nu(E))=0$ at various magnetic  fields. Left negative, the center positive, and
 the right negative parts of the curves correspond to $v^{(i)}$, $i=1,2,3$ respectively. The dimensionless value $\lambda _1 \tau_\eta $ is plotted as a function of the stationary  current $I\approx e  V_0 W v^{(i)} $.  The related values of
   the stationary electric fields $E^{(i)}$,  where $\mathcal{F}[ v^{(i)};\nu( E^{(i)} ) ]=0$,
    are presented in panels ($b$) and ($c$).
  Negative value of $\lambda _1$ corresponds to the stable solutions, while positive to the unstable one.
   ($b$,$c$): S-type current-voltage characteristics, obtained from solutions
    of equation $\mathcal{F}(v;\nu(E))=0$. The units of the current and
     the electric field are $I_{\rm Cu} = \pi^2 e n_0 W^2/64\sqrt{\tau_\Delta \tau_2}$ and $E_{\rm Cu} = \pi^2 m \eta_0/4 e W \sqrt{\tau_\Delta \tau_2} $.
  Panels ($b$) and ($c$) differ by the type of the scale in vertical axis: the logarithmic one ($b$)
  and the linear one ($c$). All curves are plotted at $\tau_\eta/\tau_1=0.09$.
}
\end{figure*}

Similarly to~\cite{Alekseev_Semina_2026}, we consider the slow varying flows,
with the frequencies $\omega$ much smaller that the relaxation
rate~$1/\tau_{2}$. In hydrodynamic regime shear stress relaxation dominates over the bulk momentum relaxation $\tau_1 \gg \tau_2$
 and thus the Ohmic component of the flow controlled by $1/\tau_1$ is relatively small (at weak magnetic fields $\omega_c \tau_2 \lesssim 1$)~\cite{je_visc}.
Therefore
we omit the time derivatives  of $\hat{\Pi}$
in the second and third equations of~(\ref{main_eq_gen}) which reduce to the following relation between $\Pi_{xy}(y,t)$
and~$\partial V (y,t)/ \partial y  $:  $  \:   \Pi_{xy}  \,
=\,    - \, m\, \eta _{xx}^{nl}  [V] \: \partial V/ \partial y  \: , $
where we have introduced the non-linear viscosity
coefficient~$ \eta _{xx}^{nl}  [V] (y,t)$:
\begin{equation}
\label{eta_nl}
 \begin{array}{c}
  \displaystyle
 \eta _{xx}^{nl}  [V]  = \frac{\eta_0 }{ \displaystyle
   g +
    \displaystyle  (2\omega_c\tau_{2}) ^2 /g }
   \:,
   \\
   \\
   \displaystyle
   g = 1 + \tau_{2}
    \tau _\Delta  \Big( \frac{\partial V }{ \partial y} \Big) ^2
   \,.
   \end{array}
\end{equation}
The resulting hydrodynamic equation for slow flows takes the form~\cite{f2}:
\begin{equation}
\label{motion_eq_fin}
   \frac{ \partial V }{ \partial t }
    \, = \,
   \frac{eE(t)}{m}
   \,   + \,
   \frac{\partial}{\partial y} \Big( \, \eta _{xx}^{nl}  [V] \,
     \frac{ \partial V}{ \partial y } \, \Big)
     \, - \, \frac{V}{\tau_1}
     \, .
\end{equation}
This closed form non-linear equation describes evolution of non-uniform hydrodynamic velocity $V(y,t)$ under the applied field $E(t)$ which vary with characteristic timescale much greater than $\tau_2$.

 The dependence of the stress component $\sigma_{xy}  =
m\, \eta _{xx}^{nl}  [V] \, \partial V/ \partial y $,
corresponding to Eq.~(\ref{eta_nl})  on the velocity
gradient $ \dot{\gamma} =\partial V/\partial  y $
is illustrated in Fig.~1. The effect of local heating on the type of non-Newtonian behavior of electron fluid varies with the magnitudes of velocity gradient $\dot{\gamma}$ and magnetic field.
In particular, (i) at classically weak magnetic fields  $ \omega_c \tau_2 <1 $ the fluid is pseudo-plastic $d\sigma_{xy} /d \dot{\gamma} < \eta_{xx}  $ and, correspondingly, $\eta_{xx}^{nl}$ decreases with
 $\dot{\gamma}$; (ii) at strong magnetic fields $ \omega_c \tau_2 >1 $ and at relatively small $\dot{\gamma} $ it becomes dilatant $d\sigma_{xy} /d\dot{\gamma} > \eta_{xx} $  
 and the non-linear viscosity $\eta_{xx}^{nl}$ increases with $ \dot{\gamma}$ while for large $\dot{\gamma}$ the pseudo-plastic behavior is reestablished. Moreover, we see that the non-linear viscosity $\eta_{xx}^{nl}$  is always non-monotonous with 
 $ \dot{\gamma} $. We will see that instabilities are also present in our system.
 
One should impose some boundary conditions on the velocity $V(y,t)$
at the sample edges, $y=\pm W/2$, in order to
 solve equation~(\ref{motion_eq_fin}).
For simplicity, we chose the no-slip boundary
conditions corresponding
 to the very rough edges: $ V|_{y=\pm W/2} = 0 $~\cite{Afanasiev_at_al_2025}.

\begin{figure*}[t!]
\centerline{\includegraphics[width=.99 \linewidth]{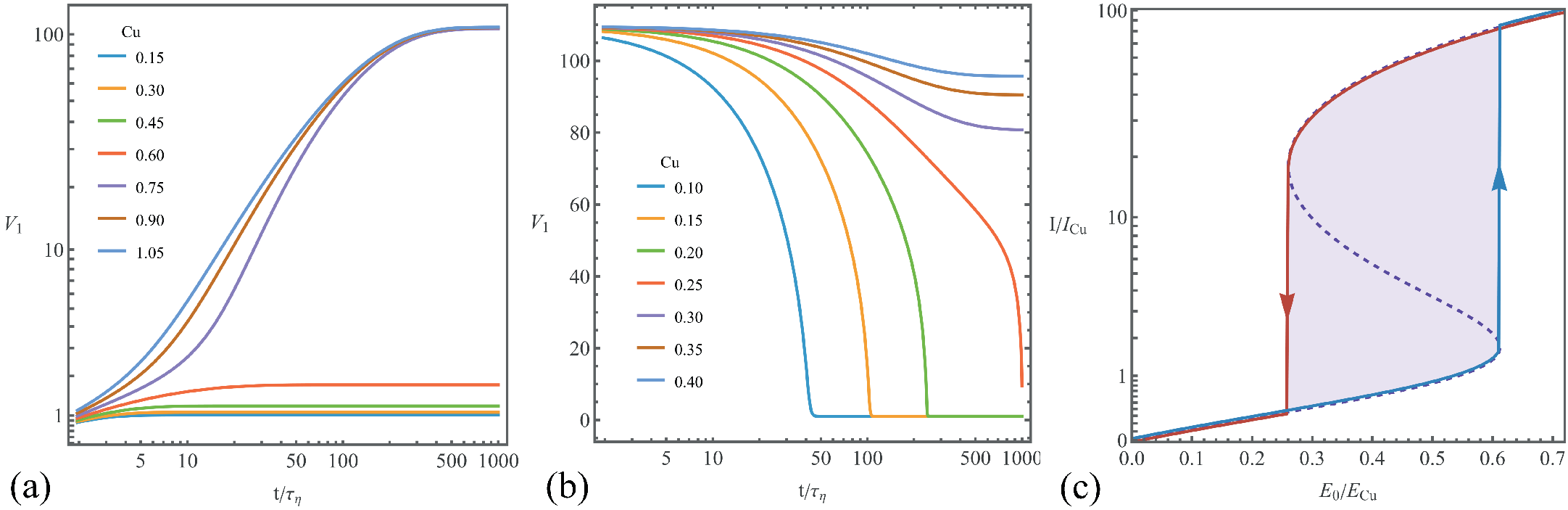}}
\caption{
($a$): Time evolution of velocity $v(t)$ for the turn on setup with the zero initial velocity $v(t=0) =0$ and 
 the electric field switched on at $t=0$ to nonzero different values $E_f$ below and above 
 the threshold fields [shown by blue line in the panel(c)].
 The magnitudes of the electric field $E_f$ 
  are presented  via the parameter  ${\rm Cu} = 4 e E_f W \sqrt{\tau_\Delta \tau_2}/\pi^2 m \eta_0 $);
($b$):
Time evolution of velocity $v(t)$
when the starting velocity  and  electric field were  
above the threshold values, corresponding to the right side of the hysteresis curve [see panel (c)],
 and then at  $t=0$
the electric field $E$ was switched down to different values $E_f$
 above and below the threshold field [shown by red line in the panel(c)];
($c$):
The resulting hysteresis curve:
the lower branch corresponds to evolution of the steady-state of $v(t)$ on panel ($a$) $E_f$, while
the upper branch corresponds to variation of the steady-state of $v(t)$ on panel  ($b$). All curves are plotted at $\tau_\eta/\tau_1=0.09$. 
}
\end{figure*}


 {\em 3. Approximate  discrete model of the fluid flow.}
Let us introduce the dimensionless variables $\widetilde{y} = y/W$
and $\widetilde{V} = V/ V_0$,
where $V_0 = eE W^2 / (m\eta_0) $ is
the characteristic magnitude of the hydrodynamic velocity
in the linear (by $E$) stationary Poiseuille flow at zero magnetic field.
 Equation~(\ref{motion_eq_fin}) in scaled variables takes the form:
\begin{equation}
\label{motion_eq_fin2}
\begin{array}{c}
  \displaystyle
   \frac{ \displaystyle  \partial  \widetilde{V} }{ \partial   \widetilde{t} }
   =
  1
   +
   \frac{ \displaystyle
   \partial    }{ \partial   \widetilde{y} }
    \, \Big( \,
   \frac{ \displaystyle
   \partial   \widetilde{V} / \partial   \widetilde{y} }
   { \displaystyle
  g +
   \beta ^2/ g }
   \, \Big) \,
   - \frac{ \displaystyle  \widetilde{V} }
   { \displaystyle   \tau_1 / \tau_\eta }
   \, ,
   \\
   \\
     \displaystyle
   \;\;
    g =   1 + \nu' \,   \Big( \frac{\partial  \widetilde{V} }
     { \partial   \widetilde{y}}  \Big) ^2
  \,
   ,
   \end{array}
\end{equation}
where $ \widetilde{t}=t/\tau_\eta $, $\tau_ \eta = W^2 /\eta_0$ is the characteristic time
of momentum relaxation in a Poiseuille flow at the channel boundaries;
$\nu' (\beta)  \,= \,  C_\nu \, f(\beta) \, \nu $ is the dimensionless
parameter of non-linearity, and its magnetic field dependence is:
\begin{equation}
\label{f_,_nu}
  f(\beta)  =  \frac{ 1 }{ 1+ \beta^2}
 \, , \quad \;\;
  \nu  =   \frac{ \varrho \, \hbar \, a^3 v_F^2 }{E_D^2 \, \tau_{2} }
    \, \Big( \, \frac{ eEW }{ m v_F^2 } \, \Big)^2
 \,.
\end{equation}
Here $ \nu  \sim E^2  $ is the magnetic-field-independent parameter
of non-linearity for the heating mechanism and~$ C_\nu $ is a numerical coefficient, see Ref.~\cite{Alekseev_Semina_2026} for details,
  $E_D$  is the deformation potential for acoustic phonons,  $a$ is the quantum well width and $\rho$ is the crystal density.

In Ref.~\cite{Alekseev_Semina_2026}, exact steady-state solutions of Eq.~(\ref{motion_eq_fin2}) were studied for moderate nonlinearity parameters when the solution is unique and smooth. In this paper, we study the regime of large-amplitude solutions $V(y,t)$,
when for some range of parameters there are two coexisting stable stationary solutions $V_{stat}(y)$, and the time evolution of $V(y,t)$ undergoes transitions between them.

A general solution to this problem can be obtained by expanding $V(y,t)$ into functional series by the complete and orthogonal basis of spatial harmonics in $y$ with time-dependent coefficients:
$V(y,t) = \sum _ n V_n(t) Y_n(y) $, where $Y_n(y)$ are some basis functions corresponding the chosen boundary conditions on $V(y,t)$. Substituting this expansion into equation~(\ref{motion_eq_fin2}) yields a system of differential nonlinear equations for the expansion coefficients $V_n(t)$.

The simplest version of this method involves truncating series expansion and leaving only the fundamental harmonic $Y_1(y)$ in it. We use the fundamental spatial harmonic of the homogeneous problem for Newtonian fluid (i.e. $\nu\to 0$ limit) fot the chosen geometry $Y_1(y) = \cos(\pi y / W)$. In this case equation~(\ref{motion_eq_fin2}) takes the simple form
\begin{equation}
\label{eq_int}
\begin{array}{c}
\displaystyle
  \frac{ \displaystyle  d \widetilde{V}_1 }{ d \widetilde{ t} }
    \, = \,
  1  \,   -  \, \frac{ \displaystyle  \widetilde{V}_1 }
   { \displaystyle   \tau_1 / \tau_\eta }
  \,  + \,
  \\
  \\
  \displaystyle
  \,+\,
    \int _{-1/2} ^{1/2}  d \widetilde{y}  \; Y_{1}(\widetilde{y})
    \, \Big[ \, \frac{  \widetilde{V} _1 \,  Y_{1}'(\widetilde{y}  ) }
   { \displaystyle
  g_1 (\widetilde{y}  )+
   \beta ^2/ g_1(\widetilde{y}  ) } \Big]'
     \,
   \:,
   \end{array}
\end{equation}
where $\widetilde{ t} = t/\tau_\eta$,   $ g_1(\widetilde{y}  )  =  1
+ \nu' \,  \widetilde{V} _1 ^2 \,  Y_1^2(\widetilde{y}  )$,
This is an ordinary non-linear differential equation for the velocity
amplitude $\widetilde{V}_1(t)$.

A rough interpolation for the above equation, based substitution of $Y_1(\widetilde{y}$ by its characteristic value in the integral term, can be presented in the form:
\begin{equation}
\label{eq_alg}
  \frac{ \displaystyle  d \widetilde{V}_1 }
   {  d  \widetilde{t} }
    \, = \,
  1
  \,  - \,
    \frac{  c_0 \, \widetilde{V} _1   }
   { \displaystyle
  g_1  +
   \beta ^2/ g_1  }
     \,
   - \frac{ \displaystyle  \widetilde{V}_1 }
   { \displaystyle   \tau_1 / \tau_\eta }
   \:,
\end{equation}
where  $ g_1 =  1 +  c_g \nu' \,  \widetilde{V} _1 ^2 $,  and  $c_0$, $c_g$
 are the numerical constants depending on the chosen function $Y_1(y)$.
 For the basis functions $Y_1(\widetilde{y}) = \cos(\pi \widetilde{y} )$, the direct numerical evaluation of the integral in (\ref{eq_int}) shows that the resulting
dependence of the viscosity term on the parameters $\beta$ and $\nu'$ is qualitatively identical to the interpolation, and their numerical values differ by no more than in 1.5-2 times.

In our numerical analysis (presented in Figs.2 and~3)
 we have used the exact equation~(\ref{eq_int}). However, the simplified equation~(\ref{eq_alg}) is useful
  as it yields a simple transparent picture for understanding the essence
of the stationary solutions, their  stability,  and the appearance of bistability.
 We will discuss this at a qualitative level in the next section.


{\em 4. Stationary solutions and their stability.}  Equation (\ref{eq_alg})
can we written in a general form:
\begin{equation}
 \label{eq_alg2}
  \dot{v}
     =  \mathcal{F} (\,v \,)
    \,, \;\;
    \mathcal{F} (\,v \,)  = 1-\frac{v}{\widetilde{\tau}_{1}}
    - \frac{\displaystyle  v }{ \displaystyle 1
     + \nu' v^2+  \frac{\displaystyle \beta^2 }{\displaystyle 1+ \nu' v^2 } }
    \,,
\end{equation}
where $\dot{v} = dv/d\widetilde{t}$,
 $ v \equiv \widetilde{V}_1 $ and $\mathcal{F}(v) $ is the right-hand
 side of (\ref{eq_alg}) which depends on three dimensionless
 parameters $ \widetilde{\tau}_{1} = \tau_1 / \tau_\eta $,   $\nu'(\beta)$, and $\beta$,
which are renormalized to exclude the parameters  $c_0$, $c_g$.

Provided $ \tau_1  \ll  \tau_\eta $, equation (\ref{eq_alg2})
 can have one ($v^{(1)}$ or $v^{(3)}$) or three ($v^{(1)}<v^{(2)}<v^{(3)}$) stationary solutions.
The corresponding derivatives $d\mathcal{F}(v^{(i)})/dv$ have the sign $-$ for the former case
and the signs $-$, $+$, $-$ for the latter. This indicates that solutions $v^{(1)}$ and $v^{(3)}$ are stable while $v^{(2)}$ is unstable, see also Fig.~2(a).

At $\nu' \ll 1$ the only solution $v^{(1)}$ is possible which corresponds to a hydrodynamic Poiseuille-like flow, characterized by moderate nonlinearity due to the heating effect. It is analogous to the case of absolutely defectless sample $1/\tau_1 =0$ considered in Ref.~\cite{Alekseev_Semina_2026} via an exact solution of equation (\ref{motion_eq_fin2}) for $V^{(1)}(y)$  in the stationary regime.

At  $ \tau_1 \ll \tau_\eta $ and  $\nu \sim 1 $
 the almost Ohmic solution $v^{(3)} \approx \tau_1 / \tau_\eta $
 appears alongside with the almost hydrodyanamic one $v^{(1)} \sim 1 $. As it was mentioned in~\cite{Alekseev_Semina_2026} and is seen for the form of $\mathcal{F}(v)$, at sufficiently strong non-linearity $\nu'(\beta) \gtrsim 1 $ the  hydrodynamic-like
 solution $v^{(1)}$ vanishes and only the Ohmic-like solution $ v^{(0)} \to v^{(3)} $ exists.

The described behavior of stationary solutions for various values  of $ \tau_1 /\tau_\eta $
 and $\nu' $  underlies emergence of S-type current-voltage characteristics. Indeed, within our model the net current up to a numerical factor is given by $I=en_0V_0v^{(1)} W $ or $I=en_0V_0v^{(3)} W $,
 while the voltage $EL$ ($L$ is the sample length) enters the both values $V_0$ and $\nu'$. In Fig.~2(b,c) we have plotted the calculated
 current-voltage characteristics
 at different values of the magnetic field parameter $\beta$.
It is seen that an increase in magnetic field reduces the amplitude of the S-type
 current-voltage characteristic.


{\em 5. Non-stationary solutions and bistability.}
Assuming that there are no other types of degrees of freedom and instabilities,
 let us study non-stationary solutions of Eq.~(\ref{eq_alg2}).
 Namely, we will be interested in the realization of the bistability regime
 with the two stable solutions $v^{(1)}$ and $v^{(3)}$
 within the transient processes.
We solve the Cauchy problem for a) zero initial flow $v(t=0) = 0$ and the step-like turned on electric field $E(t)=E_f\theta(t)$ at $t=0$ and b) purely Ohmic initial flow $v(t=0)=\widetilde{\tau}_1$ at very large value $E(t<0)=const$ (larger that all characteristic electric fields) steeply decreased to different values $E_f$ at $t>0$. We trace transient dynamics for various magnitudes of $E_f$ below and above the critical fields a which the multiple solutions appear and vanish. The solution of such problem $v(\widetilde{t})$ in the case (a)
   reproduces the single-solution parts of the I-V curve $I(E=E_f)$ and the lower part of the three-solution region. In the case (b) the function $v(\widetilde{t})$ coincides with the single-solution parts of the I-V curve $I(E=E_f)$ and the upper part of the three-solution region. This behavior is shown in Fig.~3.

As it was demonstrated in works~\cite{non-lin_hydr_1,non-lin_hydr_3,non-lin_hydr_4,non-lin_hydr_5,non-lin_hydr_6,Alekseev_Semina_2025,Alekseev_Semina_2026} the non-Newtonian electron fluid was apparently realized
 in high-quality GaAs quantum wells. Therefore, the predicted here effects of current hysteresis and flow bistability are to be expected to manifest in this material system. Nevertheless, owing to variety of observed 2D electron hydrodynamic effects and regimes in high-quality graphene samples, they also should be considered as a good candidates for discovery of hysteresis and bistability in a non-Newtonian regime of electron fluid.

It important to note that taking into account more harmonics $V_n(t)$ in the expansion
of the velocity field  $V(y,t)$ may lead to a more complex dynamics of solutions, for example,
 limit cycles and various chaotic solutions. They may have a significant impact
  on the studied stationary solutions and bistability and/or even dominate depending on
  particular parameters of the structure.

{\em 6. Conclusion.  }
The flow bistability  and the current hysteresis in the I-V characteristic are predicted
in hydrodynamic conductors with the 2D non-Newtonian electron fluid and weak background disorder scattering. Our findings open up the possibility of exploiting this systems as electronic switches and generators of ultra- high-frequency radiation.

{\em 7. Acknowledgement.  }
We thank A. A. Greshnov for fruitful discussions.

This work was financially supported by the Foundation
 for the Advancement of Theoretical
Physics and Mathematics BASIS (Grant No. 23-1-2-25-1).

Authors declare no conflict of interests.

The data supporting the findings in this manuscript are
available from the authors upon reasonable request.


\begin{thebibliography}{}
\bibitem{f2}
We emphasize that hydrodynamic equations~(\ref{main_eq_gen}) omit other inhomogeneous and non-linear effects,
such as kinematic term $(\mathbf{V} \cdot \nabla )\mathbf{V}$, thermoelectric effects, heating effect, controlled
by the energy balance with phonons  and  heat  transport (see, for reference, works~\cite{Volkov,Semina,Karpus_1,Yassievich}).
\\
\end{thebibliography}
\end{document}